# PE-GPT: A Physics-Informed Interactive Large Language Model for Power Converter Modulation Design

*Fanfan Lin*[1], *Junhua Liu*[2], *Xinze Li*[3], *Shuai Zhao*[4], *Bohui Zhao*[1],
*Hao Ma*[1], *Xin Zhang*[1] *and Xinyuan Liao*[5]

**Affiliations**: [1] *College of Electrical Engineering, Zhejiang University, Hangzhou 310027, China*
[2] *Forth AI Pte. Ltd., Singapore*
[3] *School of Electrical and Electronic Engineering, Nanyang Technological University, Singapore 639798*
[4] *Department of Energy Technology, Aalborg University, 9220 Aalborg East, Denmark*
[5] *School of Electronics and Information, Northwestern Polytechnical University, Xi'an 710072, China*

**Emails**: fanfanlin@intl.zju.edu.cn; j@forth.ai; xinze001@ntu.edu.sg; szh@energy.aau.dk; zhao_bh@zju.edu.cn; mahao@zju.edu.cn; zhangxin_ieee@zju.edu.cn; liaoxinyuan@mail.nwpu.edu.cn.

*Abstract*—This paper proposes PE-GPT, a custom-tailored large language model uniquely adapted for power converter modulation design. By harnessing in-context learning and specialized tiered physics-informed neural networks, PE-GPT guides users through text-based dialogues, recommending actionable modulation parameters. The effectiveness of PE-GPT is validated through a practical design case involving dual active bridge converters, supported by hardware experimentation. This research underscores the transformative potential of large language models in power converter modulation design, offering enhanced accessibility, explainability, and efficiency, thereby setting a new paradigm in the field.

## I. INTRODUCTION

Power converters serve as the backbone of renewable energy systems, effectively bridging the renewable energy sources and the power grid by converting and regulating power flow. Well-designed modulation strategies for these converters are pivotal, as they guarantee optimal efficiency and reliability in the process of energy conversion [1].

In general, there are different types of modulation strategies for power converters, such as phase shift modulation, duty ratio modulation, frequency modulation, and space vector modulation. Typically, the design of these strategies involves a multistep process where electrical engineers first understand the specific application scenarios, including controller settings, hardware configurations, and expected performance. This understanding guides the selection of a primary modulation strategy suited to the converter's needs. Subsequently, engineers optimize the design parameters within the chosen strategy to achieve the desired performance.

Recent advances in artificial intelligence (AI) have inspired more researchers to explore data-driven models for designing modulation parameters, aiming to enhance the design process's efficiency and effectiveness. However, these AI-based approaches have notable limitations. Firstly, they demand substantial interdisciplinary expertise from users, requiring a deep understanding of both power electronics and AI. This dual expertise is often challenging to acquire and apply. Secondly, the accuracy of AI-based designs hinges on the availability of extensive training data. The need for large datasets can be a significant hurdle, particularly in hardware experiments where such data may be limited or difficult to collect. Both challenges limit the practicality of AI-driven modulation design approaches, in terms of usability and data perspectives.

Interestingly, recent advancements in large language models, such as GPT-4 [2], Palm [3], and LLaMa [4], demonstrate superior performance in many domain-specific tasks through easily accessible interfaces. These models have been explored in diverse domains, such as online dispute resolution [5], airfoil design in aerodynamics [6], etc. Alongside, to tackle the data-intensive nature of AI-driven approaches, recent studies have begun integrating physics principles into neural networks. These physics-informed neural networks (PINNs) are being used to model phenomena like traffic flow [7], the health conditions of converters [8], and power flow in electrical systems [9], offering a promising avenue for data-light and explainable AI solutions in specialized fields.

Inspired by recent advancements, this paper introduces a pioneering approach for power converter modulation design, termed PE-GPT (GPT for Power Electronics). PE-GPT is a novel, physics-informed interactive large language model, designed to provide user-friendly, sequential guidance to users, while generating optimal modulation designs complete with thorough reasoning and analysis. Built on the GPT-4 foundation, PE-GPT's contributions can be delineated along two key dimensions:

- **Semantic aspect:** Specialized PE-customized in-context learning is applied to train GPT-4, significantly enhancing its utility and efficiency in power converter modulation design. This tailored training addresses the specific linguistic and technical nuances unique to power electronics, bridging a crucial gap in current AI applications in this field.
- **Physical aspect**: PE-GPT integrates two hierarchical physics-informed neural networks, focusing on switch-level and converter-level modeling, respectively. This dual-network approach not only endows PE-GPT with superior modulation design capabilities but also substantially reduces the volume of training data required, overcoming a major hurdle in conventional AI-driven methods.

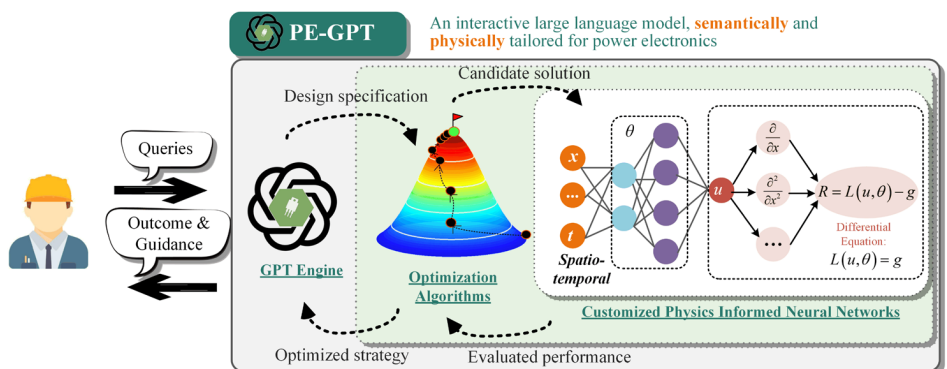

Fig. 1. Architecture of the proposed PE-GPT：a semantically and physically customized large language model for power converter modulation design.

II. THE PROPOSED PE-GPT FOR POWER CONVERTER MODULATION DESIGN

A. *The Overall Architecture of PE-GPT*

The proposed PE-GPT system, delineated in Figure 1, is an interactive large language model meticulously crafted for the power electronics domain. At its core, the architecture of PE-GPT is segmented into three interconnected components: the GPT engine, the optimization algorithm and the customized PINNs.

Users begin their interaction by inputting design queries into PE-GPT, which are then processed by the GPT Engine. This engine, leveraging domain-specific enhancements, initiates interactive dialogues to provide preliminary guidance and propose initial modulation strategy suggestions. With precise design specifications obtained by GPT engine, the optimization algorithm is activated to identify optimal modulation parameters that meet the user's requirements. This iterative optimization is refined through continuous interaction with the PINNs, which evaluate and guide the optimization toward globally optimal solutions. Finally, the GPT engine delivers the modulation strategy in a comprehensive textual format, accompanied by analytical results that substantiate the design's efficacy. With PE-GPT's synergistic architecture, it provides users not just with optimized modulation strategies but also with critical insights and explanations, facilitating a deeper understanding and much more efficient power converter design.

### B. Semantic Customization: In-Context Learning

In-context learning (ICL) [10] is one of the effective ways to adapt large language models to downstream tasks. Leveraging the ICL prompting technique, the model can be directed to tackle new tasks without the need for extensive additional training. In the context of customized power converter modulation design task, we propose a prompt template, named as *PE-Prompt*, that consists of 4 parts as the followings:

$$PE\text{-}Prompt = SMessage + SubTask + GroundC + OutputStr.$$

(1) *SMessage*: This component sets the persona that the GPT assumes during the task. For our case, we have designated the persona as "assistant to electrical engineer responsible for documenting modulation design specifications."

(2) *SubTask*: This is to list the design specifications to collect. Notably, it is important to apply chain-of-thought (CoT) technique here to instruct GPT-4 to collect specification through a step-by-step series of questions.

(3) *GroundC*: This component is to define the grounding context for the designable specifications, which includes the restrictions or ranges of the specifications.

(4) *OutputStr*: This outlines the desired format and structure of the generated output. Ensuring the alignment between the core engine and GPT-4 is crucial to enable seamless coordination without friction.

### C. Physical Customization: Two Hierarchical PINNs

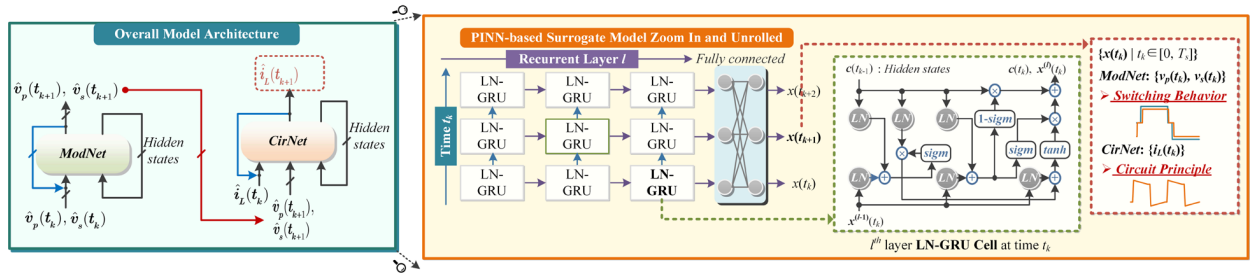

Fig. 2. The architecture of the proposed hierarchical PINNs in PE-GPT.

To ensure PE-GPT a good design outcome for the modulation strategy, it is essential to model power converters accurately. For this purpose, the proposed PE-GPT integrates two customized hierarchical PINNs, one *ModNet* for switch-level modeling to learn the switching behaviors and another one *CirNet* for the converter-level modeling to learn the circuit physics. Both *ModNet* and *CirNet* consist of several layers of gated recurrent unit with layer normalization (LN-GRU) and a fully connected layer. This paper takes the dual active bridge (DAB) converter as a studied converter. As plotted in Figure 2, $v_p$, $v_s$, and $i_L$ represent the ac voltages of the primary and secondary full bridges and the ac current through the leakage inductor.

*ModNet* infers $v_p(t_k)$ and $v_s(t_k)$ based on the information of previous time stamp $t_{k-1}$ and hidden states, catering for non-ideal switching dynamics like overshoot and oscillations. The purpose of *ModNet* is formulated in (1). Conditioned on the outputs from *ModNet*, *CirNet* recurrently predicts the inductor current $i_L$ at the next time stamp, catering for the converter-level modeling, as expressed in (2). With $i_L$ inferred from *CirNet*, diverse operating performance such as the current stress, soft switching range, and efficiency can be evaluated.

$$\left(\hat{v}_p(t_k), \hat{v}_s(t_k)\right) = ModNet\left(\hat{v}_p(t_{k-1}), \hat{v}_s(t_{k-1})\right) \quad (1)$$

$$\hat{i}_L(t_{k+1}) = CirNet\left(\hat{i}_L(t_k); ModNet\left(\hat{v}_p(t_{k-1}), \hat{v}_s(t_{k-1})\right)\right) \quad (2)$$

*ModNet* and *CirNet* are trained with two losses: the data-driven loss and the physics breaching loss. *ModNet* is trained to learn the intermediate ac voltage waveforms $v_p$ and $v_s$. *CirNet* integrates the Kirchhoff's circuit laws in (3) into its inherent feature space, where $L$, $R_L$, and $n$ denote leakage inductance, equivalent inductor resistance, and turn ratio of transformer.

$$L\frac{\partial i_L(t)}{\partial t} = -R_L i_L(t) + v_p(t) - n v_s(t) \quad (3)$$

### III. Design Case and Experimental Results

#### A. Design Case

The phase shift modulation design for the dual active bridge (DAB) converter is studied to demonstrate the proposed PE-GPT. The inference temperature of GPT-4 is set to be 0 for deterministic output. The design case using PE-GPT is demonstrated with a detailed request breakdown below and screenshots from the user interface are provided in the supplementary material.

**[Request 1] User requests to design modulation strategy of the DAB converter:** PE-GPT accepts the request and asks about the selected modulation strategy and recommends some strategies for selection.

**[Request 2] User provides the preferred modulation strategy for the design:** PE-GPT continues to guide the user to select the prioritized design objectives.

**[Request 3] User defines the prioritized design objective for the specific application:** PE-GPT continues to guide the user to provide information about the operating conditions of the application.

**[Request 4] User specifies the details of the operating conditions including power and voltage:** PE-GPT recognizes receiving the information and asks the user to select an optimization algorithm to use in the backend and recommends some options.

**[Request 5] User chooses an optimization algorithm to proceed:** PE-GPT recognizes the collection of all necessary specifications and displays the progress of design process.

**[Request 6] Display design outcomes and analysis:** PE-GPT generates the design outcomes in a visualized figure. For analysis purposes, the whole optimization landscape and the comparison of the optimized outcome with other modulation strategy are presented and explained as well.

Upon definition of all specifications, the entire design process is accomplished in less than 10 seconds, remarkably expediting the overall process. Equally importantly, PE-GPT is also able to assist in decision-making like recommending modulation strategy based on users' inputs. In this capacity, PE-GPT holds the potential to elevate the efficiency of electrical engineers and researchers.

## B. Experimental Results

The modelling accuracy of PINNs will directly affect the optimization results and thus the design outcome performance. To verify the effectiveness of the proposed hierarchical PINNs in PE-GPT, a small data size of 10 is used to train some popularly used data-driven algorithms. As shown in Figure 3, PE-GPT outperforms the second-best algorithms by 63.2%.

The practical efficacy of PE-GPT's design was verified through a hardware prototype of DAB converter under specified conditions (rated power: 200 W, input voltage: 200 V, output voltage: 160 V), as shown in Figure 4. The operational waveforms confirmed the successful implementation of the triple phase shift (TPS) modulation, achieving complete zero voltage switching. The measured peak-to-peak current stress $i_{pp}$ of 7.661 A marks a notable improvement over traditional manually designed TPS at 8.12 A and single phase shift at 9.17 A.

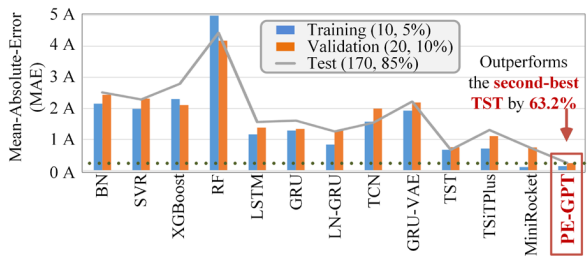
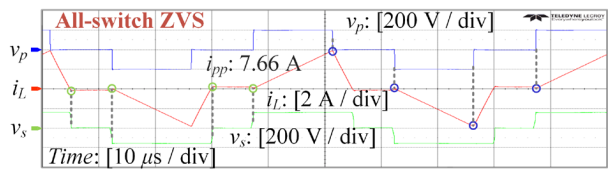

Fig. 3. Results on MAE when data size is 10.   Fig. 4. Hardware experiment waveforms for the design case.

## IV. CONCLUSION

This paper is the first attempt to in power electronics to revolutionize the design process of power converters using interactive large language model, PE-GPT. Two significant advancements have been made to adapt GPT-4 for this specialized application. First, from a semantic perspective, in-context learning is employed to tailor GPT-4, bridging the gap in domain-specific knowledge. Second, in the physical dimension, domain expertise is integrated into dual physics-informed neural networks. In a case study involving a dual active bridge converter, PE-GPT not only expedited the design process to under 10 seconds but also significantly enhanced accuracy by 63.2% compared to conventional data-driven methods, as validated by hardware experiments.

With this work, we aspire to pave the way for future research in holistic life cycle management of power converters, encompassing design, maintenance, and fault diagnosis and more. We envisage leveraging the capabilities of large language models to unlock new potentials in this field.